# Interaction of Magnetization and Heat Dynamics for Pulsed Domain Wall Movement with Joule Heating

Serban Lepadatu[*]

*Jeremiah Horrocks Institute for Mathematics, Physics and Astronomy, University of Central Lancashire, Preston PR1 2HE, U.K.*

## Abstract

Pulsed domain wall movement is studied here in $Ni_{80}Fe_{20}$ nanowires on $SiO_2$, using a fully integrated electrostatic, thermoelectric, and micromagnetics solver based on the Landau-Lifshitz-Bloch equation, including Joule heating, anisotropic magneto-resistance, and Oersted field contributions. During the applied pulse the anisotropic magneto-resistance of the domain wall generates a dynamic heat gradient which increases the current-driven velocity by up to 15%. Using a temperature-dependent conductivity significant differences are found between the constant voltage-pulsed and constant current-pulsed domain wall movement: constant voltage pulses are shown to be more efficient at displacing domain walls whilst minimizing the increase in temperature, with the total domain wall displacement achieved over a fixed pulse duration having a maximum with respect to the driving pulse strength.

[*] SLepadatu@uclan.ac.uk

# I. Introduction

The manipulation of magnetic domain walls in nanodevices has attracted continued interest due to the potential applications for magnetic memory [1] and logic [2]. It is well known that spin-polarized currents can move domain walls because of the spin-transfer torques (STT) exerted on the magnetization [3-6]. Joule heating is inevitably associated with the large current densities required to move domain walls in nanowires, which can result in changes of the nanowire resistance [7], transformations of domain wall structure [8] as the temperature approaches the Curie point of the magnetic material, as well as changes in domain wall velocities [9]. A full numerical treatment of the effect of an electrical current on the magnetization is surprisingly difficult. The applied current not only affects the magnetization directly through STT, but also generates an Oersted field which interacts with the magnetization, and generates Joule heating depending on the magnetic material, its geometry, temperature-dependent conductivity, and substrate material. Thus the electrical current also affects the magnetization indirectly, since the equilibrium magnetization, damping, and exchange stiffness values are temperature dependent. Moreover the magnetization itself modifies the current density, and therefore also the Joule heating, through its anisotropic magneto-resistance (AMR). Thermal gradients in magnetic structures can also generate domain wall motion due to the magnonic spin Seebeck effect [10-12], emphasizing the need to include the interaction between magnetization and heat dynamics in analyses of experimental results. Here it is shown that the AMR contribution of a moving domain wall generates a dynamic heat gradient which can significantly affect the wall velocity. Moreover, significant differences between constant voltage and constant current pulses are found, with constant voltage pulses resulting in less severe temperature increase for the same domain wall displacement; the total domain wall displacement achieved over a fixed pulse duration also shows a maximum with respect to the driving pulse strength. Micromagnetics studies including Joule heating effects have been published [13-15], but typically a constant current density is used throughout the simulations, either calculated analytically or imported from an external modelling software for non-rectangular geometries. Here the electrostatic, thermoelectric, and micromagnetics equations are fully integrated within the same model, allowing a detailed insight into the rich physics of the interplay between magnetization and heat dynamics, within the wider spin caloritronics field [16].

The paper is organized as follows. In Section II a method is introduced to accurately model the effect of a substrate on the temperature in the nanowire. The electrostatic solver used to compute the current density for a temperature-dependent, and spatially varying conductivity, including AMR contributions, is described; this is fully integrated with the heat flow equation solver, allowing for Joule heating effects to be accurately described. If Joule heating effects are considerable the resistance of the nanowire during a constant voltage pulse changes significantly, and thus the current density during the pulse also changes, resulting in different current-induced domain wall movement (CIDWM) behaviour compared to the constant current scenario. This is studied in Section III using the Landau-Lifshitz-Bloch (LLB) equation. The effect of AMR on the domain wall velocity, in the presence of Joule heating, is studied in Section IV.

## II. Joule Heating Modelling

For typical current densities used in CIDWM experiments Joule heating effects can be significant [7,8,17]. This is more pronounced for materials with relatively low electrical conductivity such as $Ni_{80}Fe_{20}$, and substrates with poor thermal diffusivity such as $SiO_2$, both commonly used in CIDWM experiments. The generated heat energy density due to Joule heating is given in Eq. (1), where **J** is the current density and $\sigma$ is the electrical conductivity of the nanowire.

$$Q = \frac{\mathbf{J}^2}{\sigma} \quad (\text{W/m}^3) \tag{1}$$

The heat flow is governed by Eq. (2), where $T(\mathbf{r}, t)$ is the temperature function, $C$ is the specific heat capacity, $\rho$ is the mass density, and $K$ is the thermal conductivity [18].

$$C\rho \frac{\partial T(\mathbf{r},t)}{\partial t} = \nabla \cdot K \nabla T(\mathbf{r},t) + Q(\mathbf{r},t) \tag{2}$$

The geometry we are interested in consists of a long $Ni_{80}Fe_{20}$ nanowire on a $SiO_2$ substrate, with parameters given in Table 1. Analytical formulas for such a geometry, describing the temperature change in the nanowire due to Joule heating, have been derived by You et al. [19]. This is given in Eq. (3), where $w$ and $h$ are the nanowire width and height, $C_S$, $\rho_S$, and $K_S$ are the substrate material parameters, and $\sigma_0$ is the electrical conductivity.

$$\Delta T(t) = \frac{whJ^2}{\pi K_S \sigma_0} \sinh^{-1}\left(\frac{4\sqrt{tK_S/\rho_S C_s}}{w}\right) \quad (\text{K}) \tag{3}$$

Whilst Eq. (3) is a useful formula, it does depend on a few limiting assumptions, namely constant temperature in the nanowire, temperature-independent conductivity and current density, lack of specific heat capacity for the nanowire, and a Gaussian profile for the Joule power density distribution along the nanowire cross-section. For realistic modelling of Joule heating effects we need to consider the temperature dependence of the conductivity, and must also make a distinction between constant voltage-pulsed and constant current-pulsed experiments. The electrical resistivity for $Ni_{80}Fe_{20}$ above room temperature follows a linear dependence on temperature to a good approximation, and thus the electrical conductivity is written as:

$$\sigma = \frac{\sigma_0}{1+\alpha_T(T-T_R)} \quad (S/m), \tag{4}$$

where $\alpha_T$ is the thermal coefficient, and $\sigma_0$ is the conductivity at the reference temperature $T_R$. The effect of this temperature dependence has been experimentally investigated in voltage-pulsed $Ni_{80}Fe_{20}$ nanowires and shown to be significant, with large changes in resistance recorded on time scales ranging from 50 ns – 100 ns [17,20]. In this work we will consider two scenarios: i) constant current pulses, and ii) constant voltage pulses where the voltage is applied directly across the nanowire. Fast pulse sources capable of delivering either a constant current pulse or constant voltage pulse are available, although in typical experiments the impedance of the waveguide used to deliver the pulse should be given consideration, since the voltage across the nanowire can change as the sample resistance increases with temperature [7,17]; this effect is negligible if there is a large initial impedance mismatch, and here it is useful to consider the two limiting scenarios outlined.

First, the current density may be calculated using $\mathbf{J} = \sigma\mathbf{E}$, where $\mathbf{E}$ is the electric field, $\mathbf{E} = -\nabla V$, and the potential $V(\mathbf{r})$ is obtained by solving the Poisson equation [21]:

$$\nabla^2 V = \mathbf{E}.\nabla\sigma/\sigma \tag{5}$$

The above equation is obtained for the steady state where the current continuity relation is $\nabla.\mathbf{J} = -\partial\rho_C/\partial t = 0$, with $\rho_C$ being the unpaired volume charge density; this is justified since the charge relaxation time in metals is much smaller than the micromagnetics time scales studied here. For constant conductivity Eq. (5) reduces to the usual Laplace equation, however $\sigma$ can vary spatially due to its temperature dependence and inclusion of AMR [22], Eq. (6), where $\mathbf{e}$ and $\mathbf{m}$ are the normalized electric field and magnetization respectively, and $r_{AMR}$ is an AMR ratio obtained experimentally as described in Ref. [23].

$$\sigma = \frac{\sigma_\perp}{1 + r_{AMR}(\mathbf{e}.\mathbf{m})^2} \quad (S/m) \tag{6}$$

These effects have been fully integrated into the finite difference Boris micromagnetics software, implementation details described in [24] Methods section, with Poisson's equation solved using the parallel successive over-relaxation algorithm with a relaxation constant of 1.9 for 3D simulations [25]. The potential distribution is solved initially and updated at runtime as required in order to maintain the set convergence condition $\max|\Delta V|<10^{-7}$, i.e. the maximum change in voltage (normalized using an inverse-symmetric potential drop between the two electrodes in order to minimize floating point errors) from one iteration to another in any one cell must be below the set Laplace convergence constant. The Oersted field is calculated from the solved current density using the formulas derived in Ref. [26], and included in the effective field of the micromagnetics model. The Oersted field is updated during the simulation when the current density distribution changes above a pre-set threshold. All the modules used, including the micromagnetics solvers, have been tested on the GPU using CUDA routines in both single and double floating point precision, as well as on the CPU using double floating point precision, with virtually identical results. The results presented here have subsequently been obtained using CUDA computations with single floating point precision.

Next, the effect of the substrate on the temperature inside the nanowire must be modelled. Eq. (2) is solved inside the main micromagnetics mesh, with the micromagnetics discretization cellsize, using the forward-time centred-space method (FTCS) [27] with a time sub-step typically smaller than that used for the LLG or LLB evaluation [28]. The simplest approach to modelling the effect of the substrate consists of introducing Dirichlet-type boundary conditions [27]:

$$T_B = T_0 + \alpha_B (T_W - T_0), \tag{7}$$

where $T_0$ is the base temperature ($T_0 = 293$ K), $T_W$ is the temperature inside the nanowire, and $T_B$ is the boundary temperature used when computing the differentials in Eq. (2). $\alpha_B$ is a fitting constant, $\alpha_B \in [0,1]$, with $\alpha_B = 1$ resulting in an insulating boundary. Whilst this method is simple it is only able to reasonably reproduce temperature time-dependence on very short time scales, typically up to a few nanoseconds. A related approach has been considered by Moretti et al. [14], using a Newton-type term with a fitting factor. Ideally the substrate would be fully included in the calculations, however this is computationally very expensive, in particular for the finite difference scheme, due to the large size mismatch between the magnetic nanowire and substrate. Here it is shown a good compromise may be reached by including in the computations only a small part of the substrate around the nanowire, where generally the longer the simulation is required to remain accurate during a heating or cooling cycle, the larger the substrate that is included in the simulation must be. Other methods could be used, such as prescribed boundary heat flux, or time-dependent boundary conditions, but the appeal of this approach is its general applicability, allowing the effect of any substrate to be accurately simulated on micromagnetics time-scales by simply specifying the thermal parameters with relatively small computational cost.

Here we only consider heat dissipation through the substrate since for the nanowires with small surface area studied here the heat dissipation through the substrate is much larger compared to convective heat transfer to air. The electrical contacts are considered to be sufficiently far away from the area of interest that heat dissipation through them is also neglected. The thermal conductivity also has a temperature dependence which can be experimentally determined. This was considered for both the substrate and nanowire but found to have a negligible effect on Joule heating for the materials and geometries studied here, thus the results presented are for constant values of thermal conductivity (Table 1).

**Table 1** – Thermal and electrical parameters for $Ni_{80}Fe_{20}$ and $SiO_2$ [29,30]

|  | $K$ (W/mK) | $C$ (J/kgK) | $\rho$ (kg/m$^3$) | $\sigma_0$ (S/m) at 293 K | $\alpha_T$ (K$^{-1}$) |
|---|---|---|---|---|---|
| $Ni_{80}Fe_{20}$ | **46.4** | **430** | **8740** | **1.7×10$^6$** | **0.003** |
| $SiO_2$ | **1.4** | **730** | **2200** | - | - |

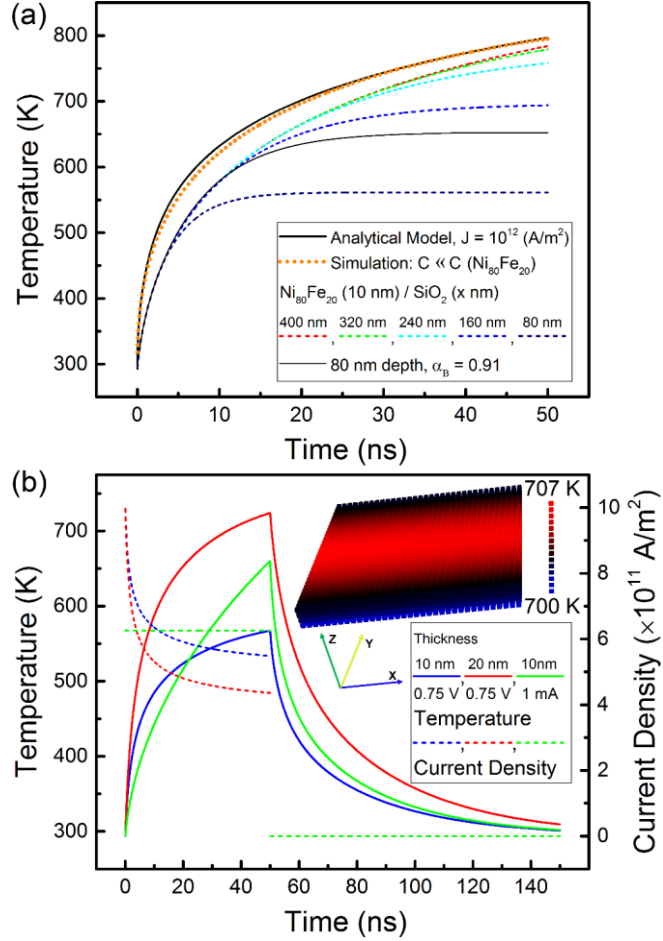

FIG. 1 (Color Online) Average nanowire temperature for 160 nm wide $Ni_{80}Fe_{20}$ nanowire on $SiO_2$ as a function of time. (a) 10 nm nanowire thickness, fixed current density of $10^{12}$ A/m$^2$, and fixed electrical conductivity. Dashed lines show simulations for different substrate depths with $\alpha_B = 0.5$. The dotted line is a simulation for a $Ni_{80}Fe_{20}$-like nanowire on 320 nm deep $SiO_2$ substrate, but with negligible specific heat capacity, to be compared with the analytical model of Eq. (3) (thick solid line). The thin solid line is a simulation for $Ni_{80}Fe_{20}$ on 80 nm deep $SiO_2$ substrate, where $\alpha_B = 0.91$ is obtained by curve-fitting to extend the duration of temperature evolution validity. (b) Average nanowire temperature (solid lines) simulated for constant voltage pulses and constant current pulse of 50 ns duration for 10 nm and 20 nm thick $Ni_{80}Fe_{20}$ as indicated in the legend, also showing the average current density (dashed lines) as a function of time. The inset shows a snapshot of the temperature distribution in the 20 nm thick nanowire.

The boundary condition in Eq. (7) is now applied to the substrate, and the interface between the substrate and nanowire, which has a discontinuity in thermal conductivity, is treated by requiring both the thermal flux and temperature to be continuous across the interface (perfect thermal contact is assumed); for full implementation details see [27]. Simulations showing the temperature change in response to a $10^{12}$ A/m$^2$ current density for a 160 nm wide, 10 nm thick Ni$_{80}$Fe$_{20}$ nanowire on a SiO$_2$ substrate, are shown in Fig. 1a, where the conductivity is fixed ($\sigma = \sigma_0$). The length of the simulated nanowire was set to 1.28 μm, but an effectively infinite nanowire is obtained by setting insulating boundary conditions at the x-axis ends (see the inset to Fig. 1b); the substrate is extended only along the y and z directions equally, with the x-axis boundaries of the simulated substrate also set as insulating, since heat flows only along the y and z directions for an infinitely long nanowire along the x-axis. Boundary conditions for Eq. (5) include fixed voltages on the right-side (ground electrode) and left-side of the nanowire; the resistance of the simulated section is ~469 Ω as expected, and a set potential of 0.75 V results in a current density of $10^{12}$ A/m$^2$ at $T = T_0$. As can be seen in Fig. 1a, increasing the depth of the simulated substrate results in an increase in the duration for which the temperature change in the nanowire is correctly reproduced – as the heat front reaches the simulated substrate boundary thermal equilibrium is quickly reached. As a rough rule the duration of validity is given by $d^2/2\mu_{th}$, where $\mu_{th}$ (m$^2$/s) is the thermal diffusivity and $d$ is the depth of the simulated substrate. The simulations may be compared to the prediction of the analytical model in Eq. (3). The most important difference is the lack of specific heat capacity for the nanowire in the analytical model, resulting in faster initial heating compared to the simulations. As a test, a good match with the analytical model may be obtained by repeating the simulation with a negligible specific heat capacity. This is shown in Fig. 1a for $C = 10$ J/kgK, noting that simulations with small values of $C$ become increasingly more difficult due to the small time-steps required and increasing floating point errors. A note on the boundary constant $\alpha_B$ in Eq. (7) used for the substrate is required: this has negligible influence on the temperature evolution, both on heating and cooling cycles, on time-scales before the heat front reaches the simulated substrate boundary – for the simulations in Fig. 1a $\alpha_B$ was set to 0.5. However, for longer time-scales $\alpha_B$ does have an effect and it may be adjusted to extend the duration of temperature evolution validity: see Fig. 1a for the 80 nm substrate depth.

Here only $SiO_2$ substrates have been discussed; in practice such substrates consist of $Si/SiO_2$, where a layer of $SiO_2$, with thickness values reaching up to 400 nm [17], is used to provide good electrical insulation from the Si wafer. Such bilayers are easily introduced into the framework developed here, but in order to simplify the analysis only the $SiO_2$ layer is modelled for the pulsed domain wall movement study. The effect of the Si substrate, with its much higher thermal diffusivity (2 orders of magnitude higher compared to $SiO_2$), is to limit the temperature increase once the heat front reaches it. Simulations using a $Si/SiO_2$ bilayer substrate with varying thickness values of $SiO_2$ are similar to those in Fig. 1a, showing a small temperature gradient with time instead of flattening out. No thermal contact resistance was used here in order to simplify the analysis; inclusion of thermal contact resistance (modelled by introducing a temperature discontinuity at the composite media interface [27]) results in a greater temperature increase, and should be considered depending on the particular experimental details, as shown by Ramos et al. [17]. Further modifications to the current framework are possible, including the use of time-dependent boundary conditions at the substrate boundaries to reproduce the temperature increase on much longer time-scales, or for substrates with high thermal diffusivity, as well as the use of a coarser discretization for the substrate alone – these are left for future work. For the pulsed domain wall movement simulations a 320 nm $SiO_2$ depth is used since this provides good accuracy over the 50 ns long pulse, as seen in Fig. 1a, with a small computational cost compared to the micromagnetics model.

If the conductivity is allowed to vary with temperature then a very different temperature variation with time is obtained in the two cases, constant voltage pulse and constant current

pulse – this is shown in Fig. 1b for the 10 nm thick nanowire, where the constant current density set is the average current density obtained over the 50 ns long constant voltage pulse. In the simulations the current is calculated from the total current density perpendicular to the ground electrode, and in the constant current mode the voltage is continuously adjusted in order to maintain a constant current. Note that even in the constant current case the current density is not uniform due to the temperature profile – see the inset in Fig. 1b – resulting in higher values of conductivity at the edges and close to the substrate, and therefore higher values of current density (the variation is around 1% from the centre to the edges); the temperature is lower at the edges of the wire due to the increased heat flow along the y direction of the substrate, in addition to the z direction. With a constant current the Joule heating is much more severe since the Joule power density in Eq. (1) increases with time. With a constant voltage the current density is higher initially, but rapidly decreases as the conductivity decreases with temperature, resulting in significantly less Joule heating.

III. Pulsed Domain Wall Movement

In the absence of temperature dependence of parameters ($T = 0$ K) the magnetization dynamics may be obtained by solving the Landau-Lifshitz-Gilbert equation with spin-transfer torque (LLG-STT), shown in Eq. (8) in implicit form [3,31].

$$\frac{\partial \mathbf{M}}{\partial t} = \gamma \mathbf{M} \times \mathbf{H} + \frac{\alpha}{|\mathbf{M}|} \mathbf{M} \times \frac{\partial \mathbf{M}}{\partial t} + (\mathbf{u}.\nabla)\mathbf{M} - \frac{\beta}{|\mathbf{M}|} \mathbf{M} \times (\mathbf{u}.\nabla)\mathbf{M} \qquad (8)$$

Here $\gamma = \mu_0\gamma_e$, where $\gamma_e = -ge/2m_e$ is the electron gyromagnetic ratio, noting $|\gamma| = 2.213\times10^5$ m/As, $\alpha$ is the Gilbert damping constant, $\beta$ is the non-adiabaticity constant, **M** is the magnetization, **H** is an effective field, and **u** is the spin-drift velocity, given by:

$$\mathbf{u} = \mathbf{J}\frac{P^0 g\mu_B}{2eM_S^0}\frac{1}{1+\beta^2} \quad \text{(m/s)} \tag{9}$$

Here $P^0$ is the current spin-polarization, $M_S^0$ the saturation magnetization, both at $T = 0$ K, with $P^0 = 0.4$ and $M_S^0 = 8\times10^5$ A/m for $Ni_{80}Fe_{20}$, $g$ the Landé $g$-factor, and $\mu_B$ the Bohr magneton. If the temperature is allowed to be non-zero, the magnetization length is no longer a constant, and in addition to the transverse damping torque we have a longitudinal damping torque. The magnetization dynamics are now described by the Landau-Lifshitz-Bloch equation [32], written in Eq. (10) in implicit form including the spin-transfer torque terms (LLB-STT) [33].

$$\frac{\partial \mathbf{M}}{\partial t} = \gamma\mathbf{M}\times\mathbf{H} + \frac{\tilde{\alpha}_\perp}{|\mathbf{M}|}\mathbf{M}\times\frac{\partial \mathbf{M}}{\partial t} - \frac{\gamma\tilde{\alpha}_\parallel}{|\mathbf{M}|}(\mathbf{M}.\mathbf{H})\mathbf{M} + (\mathbf{u}.\nabla)\mathbf{M} - \frac{\beta}{|\mathbf{M}|}\mathbf{M}\times(\mathbf{u}.\nabla)\mathbf{M} \tag{10}$$

In Eq. (10) we have $\tilde{\alpha}_\perp = \alpha_\perp/m$ and $\tilde{\alpha}_\parallel = \alpha_\parallel/m$, with $m$ being the magnetization length normalized to its zero temperature value. The transverse and longitudinal damping terms are related to the zero temperature value by $\alpha_\perp = \alpha(1 - T/3T_C)$, $\alpha_\parallel = 2\alpha T/3T_C$, where $T_C$ is the Curie temperature ($T_C = 870$ K for $Ni_{80}Fe_{20}$ [34]), for $T < T_C$. Note that with these notations, excepting the longitudinal damping torque, the LLB-STT equation has the same symbolic form as the LLG-STT equation. The effective field **H** contains all the usual contributions: demagnetizing field, direct exchange interaction field, external field, and in addition contains a longitudinal relaxation field ($T < T_C$) [32]:

$$\mathbf{H} = \mathbf{H}_{demag} + \mathbf{H}_{exch} + \mathbf{H}_{ext} + \left(1 - \frac{m^2}{m_e^2}\right)\frac{\mathbf{M}}{\chi_\parallel} \qquad (11)$$

Here $m_e$ is the temperature-dependent equilibrium magnetization given by [32] $m_e(T) = B[m_e 3T_C/T + \mu\mu_0 H_{ext}/k_B T]$, where $\mu$ is the atomic magnetic moment (for $Ni_{80}Fe_{20}$ $\mu \cong \mu_B$ [34]), $k_B$ is the Boltzmann constant and B is the Langevin function, $B(x) = L(x) = \coth(x) - 1/x$; a plot of $m_e(T)$ is shown in Fig. 2. The longitudinal susceptibility, $\chi_\parallel$, is given by $\chi_\parallel(T) = (\partial M_e(T)/\partial H_{ext})|_{H_{ext} = 0}$, where $M_e = m_e M_s^0$, thus we obtain $\chi_\parallel(T) = (\mu\mu_0 M_s^0/k_B T)\, B'(x) / (1 - B'(x)3Tc/T)$, where $x = m_e 3Tc/T$, and B' is the differential of the Langevin function. The exchange field is given by $\mathbf{H}_{exch} = (2A(T)/\mu_0 M_e^2)\,\nabla^2 \mathbf{M}$, where $A(T) = A_0 m_e^2(T)$ [35], $A_0$ being the zero temperature value of the exchange stiffness ($A_0 = 1.3\times 10^{-11}$ J/m for $Ni_{80}Fe_{20}$). The LLB-STT equation may be expanded to its explicit form, where $\tilde{\gamma} = \gamma/(1+\tilde{\alpha}_\perp^2)$, as:

$$\frac{\partial \mathbf{M}}{\partial t} = \tilde{\gamma}\mathbf{M}\times\mathbf{H} + \frac{\tilde{\gamma}\tilde{\alpha}_\perp}{|\mathbf{M}|}\mathbf{M}\times(\mathbf{M}\times\mathbf{H}) - \frac{\gamma\tilde{\alpha}_\parallel}{|\mathbf{M}|}(\mathbf{M}.\mathbf{H})\mathbf{M} + \\ \frac{1}{(1+\tilde{\alpha}_\perp^2)}\left[(1+\tilde{\alpha}_\perp\beta)(\mathbf{u}.\nabla)\mathbf{M} - \frac{(\beta-\tilde{\alpha}_\perp)}{|\mathbf{M}|}\mathbf{M}\times(\mathbf{u}.\nabla)\mathbf{M} - \frac{\tilde{\alpha}_\perp(\beta-\tilde{\alpha}_\perp)}{|\mathbf{M}|^2}(\mathbf{M}.(\mathbf{u}.\nabla)\mathbf{M})\mathbf{M}\right] \qquad (12)$$

Note, the last STT term vanishes in the LLG-STT equation since $|\mathbf{M}|$ is constant, but must be kept in the LLB-STT equation.

It is known that the steady-state domain wall velocity below the Walker breakdown threshold is given by $v = (\beta/\alpha)u$ [36]. Comparing Eq. (8) with Eq. (10), we should expect that the temperature-dependent domain wall velocity below the Walker breakdown threshold is given by:

$$v = \frac{\beta}{\tilde{\alpha}_\perp} u \quad \text{(m/s)} \tag{13}$$

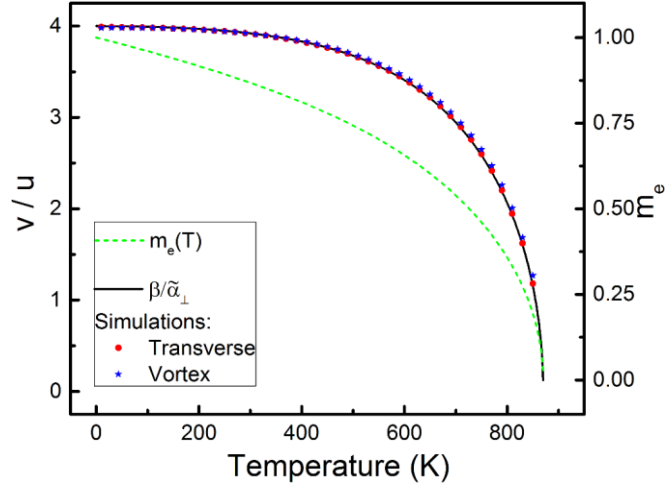

FIG. 2 (Color Online) Equilibrium magnetization function and steady-state domain wall to spin-drift velocity ratio without Joule heating, for u = 29 m/s (J = $10^{12}$ A/m$^2$), and $\beta/\alpha$ = 4. The velocities ratios were obtained from simulations for transverse and vortex domain walls and compared to Eq. (13).

The Walker threshold does vary with temperature [33], however taking a typical experimental current density of $10^{12}$ A/m$^2$, we have u ≅ 29 m/s for $Ni_{80}Fe_{20}$, and here the Walker breakdown threshold is not reached below $T_C$. This is shown in Fig. 2, where Eq. (13) is compared to simulations with no Joule heating included, taking $\beta$ = 0.04 and $\alpha$ = 0.01 for $Ni_{80}Fe_{20}$ [6].

Pulsed domain wall movement velocity curves calculated using the LLB-STT equation, including Joule heating but without AMR included, are shown in Fig. 3, also showing the total domain wall displacement and maximum temperature reached during the pulse. The cellsize used was 5 nm [24]; simulations with a 2.5 nm cellsize do not differ. The moving mesh algorithm described previously [24] was used, where the temperature in the nanowire and substrate have also been included in the algorithm here. The domain wall velocity was extracted from the simulated displacement as a function of time, by extracting the gradient using linear regression with a time stencil of 90 ps. Both for transverse and vortex domain walls it is found that constant voltage pulses are more efficient at displacing domain walls whilst minimizing the increase in temperature. With constant voltage pulses the current density starts at a high value, providing a strong initial boost, but drops quickly as the sample temperature increases (see Fig. 1b), thus reducing the temperature increase rate. With constant current the initial temperature increase is slower but continues to increase steadily throughout the pulse, slowing the domain wall velocity the longer the pulse is kept. For example comparing the -0.75 V and -1 mA pulses for the transverse domain wall in Fig. 3a, the distance covered is roughly the same (3.1 μm) but the temperature increase is ~90 K greater for the current pulse. Longer pulses result in a greater discrepancy, as the much faster temperature increase for constant current pulses on longer time scales result in drastically reduced domain wall velocities (see Fig. 2). Note that for this example the current density for the -1 mA constant current pulse is roughly the average value obtained for the -0.75 V constant voltage pulse (see the dashed lines in Fig. 3 showing the spin-drift velocity which can be converted to current density using Eq. (9)). The same conclusion holds for the vortex domain wall in Fig. 3b, where the -0.45 V and -1.3 mA pulses displace the domain wall by the same distance (2.3 μm) but with a greater increase in temperature for the constant current pulse.

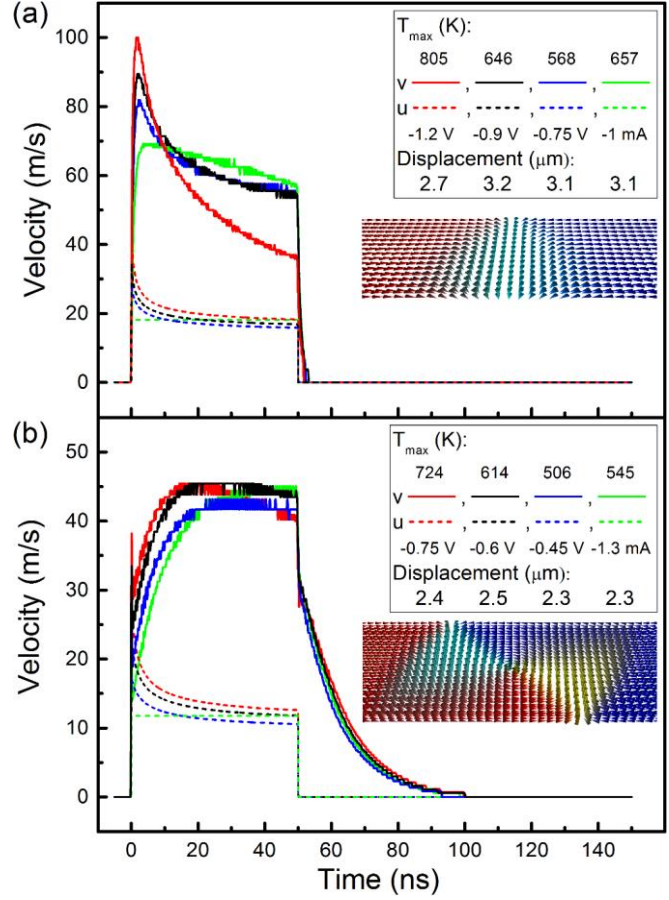

FIG. 3 (Color Online) Domain wall velocity simulated using the LLB-STT equation for 160 nm wide $Ni_{80}Fe_{20}$ nanowires on $SiO_2$, where $\beta/\alpha = 4$. The solid lines represent the domain wall velocity and the dashed lines represent the spin-drift velocity. Both voltage-pulsed and current-pulsed velocity curves are shown, as indicated in the legend, also showing the total domain wall displacement and maximum temperature reached. (a) 10 nm thick, transverse domain wall (inset), with resistance between contacts of ~469 Ω, and (b) 20 nm thick, vortex domain wall (inset), with resistance between contacts of ~234 Ω.

Another consequence of Joule heating, and the associated decrease in domain wall velocity with temperature, is the maximum displacement that can be achieved over a fixed pulse duration has a maximum with respect to the driving pulse amplitude even before the Curie temperature is reached. This is shown in Fig. 3, where for both the transverse and vortex domain walls, the stronger voltage pulses result in smaller displacements compared to the weaker pulses; the same behaviour is obtained if the current pulse amplitude is increased. Even though the initial velocity is higher, the drastic increase in temperature quickly reduces the domain wall velocity far below that obtained with a weaker pulse. If we take the transverse domain wall case, which has a small inertia over the 50 ns long pulse, the displacement is obtained by integrating Eq. (13):

$$d_u^{TW}(\tau) \cong \frac{\beta}{\alpha} \int_0^\tau u(T_u(t)) \frac{m_e(T_u(t))}{1-T_u(t)/3T_C} dt \tag{14}$$

For example for a constant current pulse, $d_u$ on the one hand is proportional to the driving strength $u$, but as the temperature approaches $T_C$ for larger values of $u$, the non-linear decrease in $m_e$ results in a maximum $d_u$ with respect to $u$. This shows that in order to maximize the distance travelled, the temperature during the pulse should be kept well below the Curie temperature, where the decrease in $m_e$ with temperature is still approximately linear. This is even more pronounced for vortex domain walls since the significant inertia [37] reduces the initial boost experienced at lower temperatures.

## IV. AMR-Generated Dynamic Heat Gradient

It is known that domain walls can move in heat gradients, shown both theoretically [10] and experimentally [11,12]. The wall motion is always towards the hotter side, resulting mainly from an imbalance in the direct exchange field as modelled in the LLB equation, due to the temperature-dependent magnetization. In the uniform cross-section nanowires considered here no such heat gradient is generated in the model considered thus far. It is well known that magnetic materials have an AMR contribution [22], resulting in a local dependence of the electrical conductivity on the angle between the current density and magnetization, as shown in Eq. (6). Here we take a 0.02 value for the AMR ratio, measured previously in thin $Ni_{80}Fe_{20}$ films [23]; for simplicity, any dependence of the AMR ratio on temperature above $T_0$ is not considered here. Due to the change in conductivity the Joule heating is also affected [38]. In particular for $Ni_{80}Fe_{20}$, since the conductivity is highest for magnetization components transverse to the current direction, the current density is modified by the domain wall, with the longitudinal current density being the dominant component. This is shown in Fig. 4b, where the current is shunted through the base of the V-shaped transverse domain wall; the conductivity is also displayed in Fig. 4b. The Joule heating power density, Eq. (1), depends on both the current density and conductivity, and is also shown in Fig. 4b. The increase in conductivity at the domain wall dominates this term, resulting in decreased Joule heating around the centre of the wall. Thus, with the domain wall at rest a temperature trough is centred on the domain wall, with both sides experiencing equal temperature gradients; when the domain wall starts to move however, the temperature trough begins to lag due to the finite heat diffusion time, resulting in a lower temperature on the trailing side of the wall – the moving domain wall experiences a dynamically generated heat gradient. Note, at the edges of the domain wall, the Joule heating power density reaches a local maximum as seen in Fig. 4b – the temperature thus reaches a maximum value at the leading edge of the domain wall, particularly in the centre of the wire.

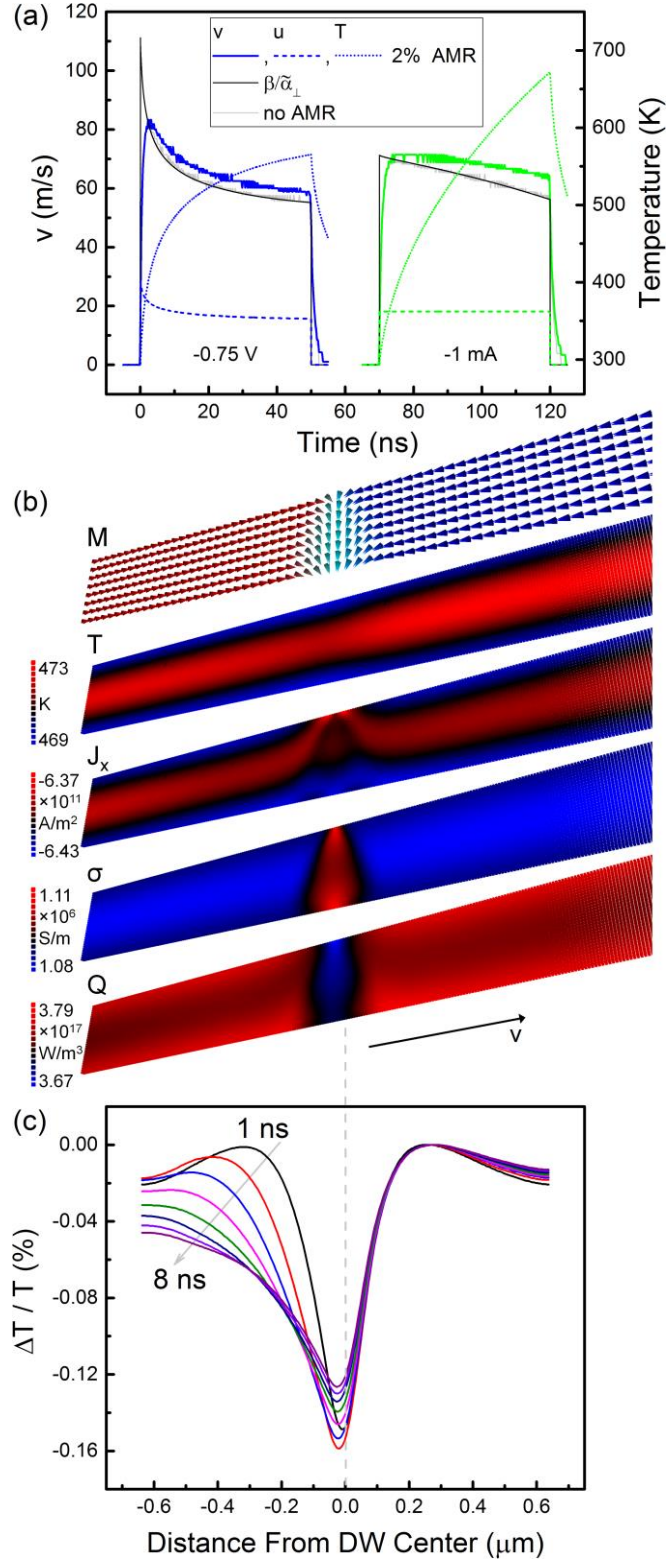

FIG. 4 (Color Online) Effect of AMR on domain wall movement for 160 nm wide, 10 nm thick $Ni_{80}Fe_{20}$ nanowire on $SiO_2$, where $\beta/\alpha = 4$. (a) Velocity for 2% AMR, for -0.75 V and -1 mA pulses (solid thick lines), including spin-drift velocity (dashed lines), and temperature (dotted lines), as a function of time. The thin black lines show the velocity computed with Eq. (13) from the simulated spin-drift velocity and temperature; this is compared with the velocity obtained from simulations without AMR (thin gray lines). (b) Snapshot of the magnetization, temperature, longitudinal component of current density, conductivity, and Joule power density at 8 ns after the start of the -0.75 V pulse. (c) Normalized longitudinal temperature profile through the middle of the nanowire, as a function of time for the first 8 ns of the -0.75 V pulse.

Domain wall velocity curves have been re-calculated for the -0.75 V and -1 mA pulses now using an AMR contribution – these are shown in Fig. 4a. The temperature and spin-drift velocity for the simulations with AMR are also shown in Fig. 4a; these are very similar to those obtained for the no AMR case – for the current pulse the temperature is only slightly higher (less than 10 K at the end of the pulse), whilst for the voltage pulse the spin-drift velocity is slightly lower (1% lower at the end of the pulse), due to the lower overall conductivity. These differences are really small however, and cannot account for the significant difference in velocities for the no AMR and AMR cases shown in Fig. 4a (~15% difference for the current pulse). To see this, Eq. (13) can be used to calculate the domain wall velocity, bearing in mind this does not take into account effects of domain wall inertia or heat gradients. The result is in very good agreement with the simulations for the no AMR case, as shown in Fig. 4a; the wall velocities calculated using Eq. (13) from the temperature and spin-drift velocity for the no AMR cases are also very similar to those shown in Fig. 4a, again showing the significant decrease in the wall velocities obtained from full simulations including AMR must be accounted for by a different mechanism. Fig. 4c shows the normalized longitudinal temperature profiles taken from the centre of the nanowire as a function of time for the first 8 ns. At the start the temperature is symmetric about the wall position, however as the wall moves the temperature profile becomes asymmetric, with the trough lagging behind the wall centre, and the temperature on the trailing side significantly lower; the domain wall thus experiences a temperature gradient which increases its velocity.

Temperature gradients also generate electrical currents due to the classical Seebeck effect. In this case, the current density becomes $\mathbf{J} = \sigma(-\nabla V - S\nabla T)$, where $S$ is the Seebeck coefficient. To include this effect in computations, Eq. (5) must be replaced with Eq. (15) which is also obtained under the current continuity condition, $\nabla \cdot \mathbf{J} = 0$.

$$\nabla^2 V = -(\nabla V + S\nabla T) \cdot \nabla\sigma/\sigma - S\nabla^2 T \tag{15}$$

Taking a value of $S = -10$ μV/K for $Ni_{80}Fe_{20}$ [39], independent of temperature for simplicity, this effect was found to be relatively negligible; the Seebeck electric field was found to be 3 orders of magnitude smaller than the externally generated electric field. Finally, the effect of the Oersted field has also been investigated, however domain wall velocities with and without the Oersted field did not show any significant differences.

## V. Summary

Here a fully integrated electrostatic, thermoelectric, and micromagnetics solver was developed, allowing a detailed study of pulsed domain wall movement in $Ni_{80}Fe_{20}$ nanowires on a $SiO_2$ substrate. A framework for accurately modelling the effect of the substrate on the temperature in the nanowire due to Joule heating was developed. It was shown that it is sufficient to model only a small portion of the substrate around the nanowire, without relying on fitting constants, where the longer a simulation is required to remain accurate, the larger the modelled substrate must be; over the 50 ns long pulses studied here the computational cost of including the substrate is small in comparison to the micromagnetics model. The use of a temperature-dependent conductivity, as obtained in experimental studies, results in significant differences between the constant voltage-pulsed and constant current-pulsed domain wall movement, for both transverse and vortex domain walls. With constant current pulses the Joule heating is more severe due to the increase in electric field with temperature required to maintain a constant current, whilst for constant voltage pulses the current density rapidly drops from its initial value, resulting in significantly decreased Joule heating; a current pulse which results in the average current density obtained over a corresponding voltage pulse was found to displace the domain wall by roughly the same amount, but result in a much higher temperature increase (over 90 K for a transverse domain wall displacement of 3.1 μm). Due to the non-linear decrease of domain wall velocity with temperature, the maximum displacement that can be achieved over a fixed pulse duration was found to have a maximum with respect to the driving pulse amplitude for both voltage and current pulses. Inclusion of AMR was found to result in a dynamically generated heat gradient which increases the domain wall velocities by up to 15%. The higher conductivity at the domain wall dominates the Joule heating power density and results in decreased Joule heating; when the domain wall moves, a positive temperature gradient is generated in the direction of motion, which acts to increase the domain wall velocity.